# ON LOCAL SPACE-TIME OF LOOP TOPOLOGICAL DEFECTS IN A NEWTONIAN LATTICE

G. Gremaud

Swiss Federal Institute of Technology of Lausanne

CH-1015 Lausanne, Switzerland

ABSTRACT

One of the most fundamental questions of modern physics is the nature of space-time. There are various propositions on the table, as the Grand Unified Theory, Quantum Gravity, Supersymmetry, String and Superstring Theories, and M-Theory. However, none of these propositions is able to consistently explain what matter is and how matter interacts by describing at the same time electromagnetism, relativity, gravitation, quantum physics and observed elementary particles.

Here it is proposed that space-time is an isotropic Newtonian 3D-lattice, and that its topological defects, namely diverse dislocation and disclination loops, are the fundamental building blocks of ordinary matter. We find, for an isotropic lattice obeying Newton's laws and with specific assumptions on its elastic properties, that the macroscopic behaviour of this lattice and of its topological defects displays "all" known physics, unifying electromagnetism, relativity, gravitation and quantum physics, and resolving some longstanding questions of modern cosmology. Moreover, studying various possible microscopic structures of the Newton's lattice, for example assuming lattice axial symmetry, represented by a "colored" cubic 3D-lattice, we show that one can possibly identify a lattice structure whose topological defect loops coincide with the complex zoology of elementary particles, opening a promising field of research.

## 1. Introduction

Since the 19[th] century, physicists have attempted to develop a unified field theory *(1)*, sometimes called the *theory of everything (ToE)*, which would consist of a single coherent theoretical framework able to account for all the fundamental forces of nature. Various theories have emerged over the past few decades, as:

- The *Grand Unified Theory (GUT) (2)*, which merges the three gauge interaction forces of the Standard Model, electromagnetic, weak and strong, into one single force,

- *Quantum Gravity (QG) (3)*, which attempts to describe the quantum properties of gravity, in order to reconcile general relativity with the principle of quantum mechanics. However, as this theory is not renormalizable, theorists have taken up more radical approaches of this problem, among those the most popular ones are *Loop Quantum Gravity (LQG) (4)* and *String Theories*,

- *Supersymmetry (SUSY) (5-10),* which proposes an extension of the space-time symmetry relating the two basic classes of elementary particles, bosons and fermions, by associating each particle from one group with a particle from the other, called its superpartner, in order to solve many mysterious features of particle physics, as well as the cosmological constant problem,

- *String and Superstring Theories (11-18)*, which are theoretical frameworks in which the point-like particles are replaced by one-dimensional objects called strings, by modelling them as vibrations of tiny strings or supersymmetric strings. String theories aim to explain all types of observed elementary particles using quantum states of these strings. In addition to the particles postulated by the standard model of particle physics, string theories naturally incorporate gravity,

- *M-Theory (19-27),* which was proposed as a unifying theory of five different versions of string theory. A surprising property of *M-theory* is that extra dimensions are required for the theory's consistency. In this regard, *M-theory* presents some analogy to the *Kaluza–Klein theory*, in which applying general relativity to a five-dimensional universe (one of the dimensions curled up) looks from the four-dimensional perspective like the usual general relativity together with Maxwell's electrodynamics.

Since the 1990s, many physicists believe that 11-dimensional *M-theory* is the *theory of everything*. However, there is clearly no widespread consensus on this issue. At present, there is in fact no candidate theory of everything that actually includes the standard model of particle physics and the general relativity. For example, no candidate theory is able to calculate the fine structure constant or the mass of the electron. Most particle physicists expect that the outcome of the ongoing experiments – the search for new particles at the large particle accelerators and the search for dark matter – are needed in order to provide further input for a theory of everything.

In this paper we consider the deformation of a Newtonian isotropic 3D-lattice, with specific elastic properties (see below), described using Euler's coordinates in an absolute reference frame. Considering the topological singularities of this *"cosmological lattice"*, namely vacancy and interstitial edge dislocation loops, mixed dislocation loops, screw disclination loops and edge disclination loops, we demonstrate that they obey to a single formalism reflecting at the same time Maxwell's equations, theory of special relativity, Newtonian gravity, general theory of relativity, and the law of quantum physics. In fact, the topological singularities of the cosmological lattice play the role of ordinary matter, interacting via the various generalized deformation field of the lattice itself.

At the microscopic level, considering the particular case of an isotropic cubic lattice with axial symmetry, we show that the set of elementary and composed topological singularities in part coincide with the known elementary particles, as described by the standard model, although the final structure is still open to question.

The complete work is accessible at the following Internet address *(28)*. It is based on an original approach of the continuum mechanics of solid lattices by Euler coordinates and of the lattice topological singularities by charge concepts, which were both developed in detail in a book published in 2013 *(29)*. In this paper, the steps of the demonstration contained in *(29)* and *(28)* are reported as briefly and succinctly as possible.

1. The Eulerian description of Newtonian lattice deformation

In ref. *(29)*, we have shown that the use of Euler's coordinates for the description of solid lattices is much more powerful than the use of conventional Lagrange's coordinates. Using vector notations of the tensors, it allows a very detailed description of the *distorsions* (deformations and rotations) and the *contorsions* (torsions and flexions) of a lattice, even in the case of very strong distorsions. Adding the physical properties of the lattice, as its Newtonian behaviour, and the first and the second principles of thermodynamics, this Eulerian theory of solid lattice deformation allows to write the complete set of equations describing its space-time behaviour, and to introduce various phenomenological properties of the lattice, as elasticity, anelasticity, plasticity, self-diffusion and structural transformations.

2. The Eulerian description of lattice topological singularities

The description of topological singularities, which can occur in a lattice as *dislocations*, *disclinations* and *dispirations*, is a domain of physics, which was initiated by Volterra's idea of macroscopic defects in 1907*(30)*. This domain has shown a very quick development during the twentieth century, as well described by Hirth *(31)*. Dislocation theory was developed in 1934 with the papers of Orowan *(32)*, Polanyi *(33)* and Taylor *(34)*, who independently introduced the edge lattice dislocation, followed by the paper of Burgers *(35)* in 1939, who described the screw and the mixed lattice dislocations. In 1956 finally, Hirsch, Horne & Whelan *(36)* and Bollmann *(37)* independently observed dislocations in metals using electron microscopes. With regard to disclinations, it is in 1904 that Lehmann *(38)* observed this kind of defect for the first time and in 1922 that Friedel *(39)* gave the first physical description of it. During the second half of the twentieth century, this domain of physics has grown considerably.

Generally, one uses differential geometries in order to describe the topological singularities of a lattice. This was initiated by the work of Nye in 1953 *(40)*, who for the first time showed that the dislocation density tensor is responsible for a curvature of the lattice, whereas Kondo in 1952 *(41)* and Bilby in 1954 *(43)* independently proved that the lattice dislocations could be identified as a crystalline version of Cartan's concept of continuum torsion *(43)*. Kröner has formalized this approach in detail in 1960 *(44)*. But the use of differential geometries rapidly becomes very complicated, due to the mathematical formulation which is similar to the formulation of general relativity, and when one has to introduce topological defects other than dislocations in the lattice. For example, Kröner had suggested in 1980 *(45)* that extrinsic point defects could be introduced as an extra-matter in the form of Einstein equations, which would lead to a purely Riemannian differential geometry without dislocations. He also proposed to introduce intrinsic point defects (vacancies, interstitials) as a non-metric part of an affine connection. Finally, he suggested to use more complicated geometries, as Finsler's or Kawaguchi's geometry, to describe topological defects other than dislocations, as for example disclinations. All these differential geometries are mathematically very difficult to manipulate, as shown for example by the mathematical dislocation theory of Zorawski published in 1967 *(46)*.

For this reason, we have developed a novel approach to lattice topological singularities *(29)*. It is based on a rigorous formulation of the concept of *"deformation charge"*, describing lattice topological singularities in Euler's coordinates: the *dislocation charges*, representing the plastic distorsions (rotations and deformations) of the lattice, and the *disclination charges*, representing the plastic contorsions (torsions and flexions) of the lattice.

These charges can only appear as *strings* or *membranes* inside the 3D-lattice. They satisfy Maxwell's equations, their energy satisfies the famous Einstein equation $E_0 = M_0 c^2$ and they present relativistic behavior. On the other hand, we prove that the long range perturbations of the lattice by localized topological singularities can be completely resumed by two vector fields and one scalar field: the *vector rotation field*, which corresponds to the electrical field, the *vector curvature field* and *the scalar expansion field*, which correspond both to the gravitational field. It is not the first time that analogies between the deformation theory and other modern physics theories are found, as shown by Kröner *(44,45),* Whittaker *(47)* and Unzicker *(48)*. But none of these analogies were as fully exploited as in *(29)*.

## 3. The cosmological lattice and its Newton's equation

In the second work *(28)* we were able to find a particular 3D-lattice, the *cosmological lattice*, containing loop topological singularities, with the following specially chosen elastic distorsion *free energy* expressed per volume unit

$$F^{def} = -K_0 \tau + K_1 \tau^2 + K_2 \sum_i (\vec{\alpha}_i^{el})^2 + 2K_3 (\vec{\omega}^{el})^2 \tag{1}$$

where $\tau$ is the *scalar volume expansion*, $\vec{\alpha}_i^{el}$ the *elastic shear strain tensor*, $\vec{\omega}^{el}$ the *elastic local rotation vector*, and $K_0, K_1, K_2, K_3$ are the *elastic modules*.

The lattice dynamics in Euler's coordinates is given by a *localized Newton's equation*

$$nm \frac{d\vec{\phi}}{dt} = -2(K_2 + K_3) \overrightarrow{\mathrm{rot}}\, \vec{\omega}^{el} + \left(\frac{4}{3} K_2 + 2K_1\right) \overrightarrow{\mathrm{grad}}\, \tau + \overrightarrow{\mathrm{grad}}\, F^{def} + 2K_2 \vec{\lambda} \tag{2}$$

where $\vec{\phi}$ is the *local lattice velocity*, $m$ the *inertial* mass of the lattice cell, $n = n_0 \, e^{-\tau}$ the *density of lattice cells*, and $\vec{\lambda}$ is the *flexion charge density* inside the lattice.

Under the following conjectures about the elastic modules

$$K_3 = K_0 > 0 \quad ; \quad 0 < K_1 \ll K_0 \quad ; \quad 0 < K_2 \ll K_0 \tag{3}$$

the lattice's Newton's equation (2) becomes the fundamental equation which allows to unify electromagnetism, relativity, gravitation, and quantum physics.

In this cosmological lattice, only *circularly polarized transversal waves* can propagate, which correspond well to light propagation by photons. When scalar expansion of the lattice is smaller than a critical value, *longitudinal waves* disappear and are replaced by *local longitudinal expansion vibrations*, which correspond to gravitational perturbations.

Transversal waves are curved by the scalar gravitational field (the local volume expansion field) due to localized topological singularities, and can even disappear in *«black holes»*.

If the cosmological lattice is finite in the absolute space, it can also present cosmological expansion and/or contraction, with all the properties described by modern cosmology, as the *Big Bang*, *inflation*, an *accelerated expansion velocity step* and even the *Big Crunch*. The origin of the *«dark energy»* postulated in astrophysics in order to explain the step of accelerated expansion velocity is now simply explained on the basis of the expansion elastic energy stored in the lattice.

## 4. Maxwell's equations and special relativity

From the rotational part of Newton's equation (2), we can deduce all Maxwell's equations of electromagnetism, including the constitutive relations between electromagnetic fields and the electrical charges and currents. In the frame of these equations, "*magnetic monopoles*" cannot exist, but new "*vector electric charges*" could.

Using Newton's equation of the cosmological lattice, we are able to calculate the elastic distorsion energy and the kinetic energy stored by topological singularities moving inside the lattice. We can then show that the loop topological singularities in this lattice satisfy the special relativity. For example, the calculation of the energies of an electron (which corresponds to an interstitial edge dislocation loop associated with a screw disclination loop) allows a simple explanation of the paradox of the electric electron energy (see *(49)* for an example). The cosmological lattice can in fact be considered as the *«aether»,* with which one can understand the dilatation of time, the contraction of length, the Michelson-Morley experiment and the Doppler-Fizeau effects, and give a very simple explanation to the twin paradox of the special relativity.

## 5. Gravitation, general relativity, cosmology and the weak interaction

With Newton's equation (2) of the cosmological lattice, we also have access to the gravitational properties of the loop topological singularities, which are more or less identical to the gravitational properties described by Einstein's gravitation theory. For example, time and lengths for a local observer, situated inside the lattice and himself constituted from topological singularities of the lattice, are affected by the local gravitational field (the scalar volume expansion of the lattice) in the same way as in general relativity. This leads to invariant Maxwell's equations for this local observer, who then perceives the light speed as a perfect constant, while the light speed strongly varies with the local scalar lattice expansion if measured by an imaginary external observer situated in the absolute space-time frame.

Some differences appear only at very short distances of topological singularities, leading to a difference with *the Schwarzschild metric* of general relativity at short distances of a singularity, and to different characteristics of the black holes radii: the radii of the photon sphere and of the Schwarzschild sphere become identical and equal to $R_{Schwarzschild} = 2GM/c^2$, when the radius where the observer's time becomes infinite disappears (the radius becomes in fact equal to zero).

It appears also that the edge dislocation loops present a curvature charge responsible for a curvature of the lattice, which can be described as a *small curvature gravitational mass*, which can be positive or negative following the interstitial or vacancy nature of the loop (corresponding respectively to matter and antimatter), and which adds to the inertial mass of the edge dislocation loops to form the total gravitational mass. Such a concept of curvature charge is completely new, as it <u>does not appear</u> in general relativity, in quantum physics or in the standard model of elementary particles. Moreover it leads to a surprising negative gravitational property *(antigravity)* of the interstitial edge dislocation loop (which corresponds to the *electron neutrino*), when all the other topological loops present normal gravitational property, even the vacancy edge dislocation loop (which corresponds to the *electron anti-neutrino*).

We show that this curvature charge is also responsible for several unexplained properties in physics, such as the *small asymmetry existing between matter and antimatter* (based respectively on interstitial and vacancy edge dislocation loops), the disappearance of the antimatter during the cosmological evolution of the universe and the «*dark matter*» which is necessary in astrophysics to explain the gravitational properties of galaxies («dark matter» corresponds here to a sea of repulsive neutrinos in which galaxies are immerged).

On the other hand, the very short distance elastic interaction between an edge dislocation loop and a screw disclination loop leads to a behaviour corresponding to the *weak interaction of the standard model* of particle physics.

Finally, a detailed description of the gravitational behaviour of the different topological singularities allows to propose a very satisfactory model for the *cosmological evolution of the matter and antimatter* inside the universe. For example, the formation of galaxies can be attributed to a *precipitation phase transition*, when the disappearance of antimatter can be attributed to a *coalescing process* of the anti-matter inside the newly formed galaxies, leading to the formation of enormous black holes in the galaxies' centres.

In the frame of this cosmological lattice, phenomena as *Hubble's expansion*, *galaxies redshift* and *cosmological background radiation cooling* also find very simple explanations.

6. The photons, the quantum physics and the particles spin

With a *quantification conjecture* $E = \hbar\omega$ *of the energy*, we begin to show that circularly polarized transversal wave packets present all the properties of *photons* (zero mass, non-zero momentum, non-locality, wave-particle duality, entanglement and decoherence).

Then, with Newton's equation of the cosmological lattice, we show that dynamical gravitational perturbations (local volume expansion fluctuations of the lattice) are always associated with moving topological singularities. The *Schrödinger equation* of quantum physics is then directly deduced from Newton's equation, giving for the first time a simple physical meaning to the quantum wave function of the topological singularities: The quantum wave function represents *the amplitude and the phase of the local gravitational fluctuations* associated to the moving topological singularities, which allows simple explanations of the probabilistic interpretation of the wave function and of Heisenberg's principle.

Applying Newton's equation (2) in the case of two coupled topological singularities, we also find simple physical explanations for the *bosons*, the *fermions*, and Pauli's exclusion principle.

Finally, we show that a static solution of this equation cannot exist in the heart of a loop topological singularity, which implies that a dynamical solution has to be found. The simplest solution is a quantified rotation of the loop around one radius, which corresponds in fact to a *quantum spin of the loop*. The arguments of the pioneers of quantum physics assuming that the equatorial velocity of a charge rotation is too large in comparison to the light speed is here swept away by the fact that the local gravitational expansion at the vicinity of the loop heart is so high that the light speed is much higher than the equatorial velocity.

7. The standard model of elementary particles and the strong interaction

All the properties described above are independent of the exact microscopic structure of the 3D-lattice, granted that the lattice be isotropic and satisfy Newton's equation (2), and the specific elastic modules obey the relations (3).

However, assuming a cubic lattice with *axial properties*, resulting in special layout and rotation properties of the lattice planes (imaginarily «colored» in red, green and blue), we can define loop topological singularities with properties similar to all the elementary particles, *leptons and quarks*, of the first family of the standard model, as well as loops similar to the *intermediate bosons*. One finds also that an interaction force acts between the loops corresponding to the quarks, due to a *"colored" lattice-stacking fault*, which presents an asymptotical behaviour corresponding to the *strong force of the standard model*. This force is also related to other «bicolor» loops presenting properties corresponding to those of the *gluons* of the standard model.

On the other hand, in order to take into account the three families of particles of the standard model, we show that a more complicated structure of the edge dislocation loops, formed as doublets of edge disclination loops, could produce at least *two supplementary families of leptons and quarks*. In fact, the real structure of the lattice is still a question which can be discussed, and other lattice structures are certainly possible, which is opening a promising field of research.

Other more hypothetical consequences have also been imagined in *(28)* in the frame of the cosmological lattice and its loop topological singularities, such as the existence of:

- *supersymmetric* topological loops,
- a *forth family of leptons and quarks* in the standard model,
- some *exotic* leptons,
- local stable macroscopic fluctuations of the gravitational field leading to a *multiverse theory* in an infinite lattice,
- stable quasi-particles formed by local microscopic fluctuations of the gravitational field, which could be called *gravitons*, but which are very different from the gravitons postulated on the basis of general relativity.

## Conclusion

It is remarkable that loop topological defects in an isotropic 3D-lattice described with Euler's coordinates in an absolute space-time frame allow to find all observed natural phenomena, by considering that
- matter is composed of those loop topological defects,
- electric charge, inertial mass and gravitational mass reflect the intimate geometry of these various topological defects,
- electromagnetism, the relativity and the gravitation reflect the long range elastic interactions between topological singularities conveyed by the lattice, which can then be considered as an "aether" for the topological defects,
- the weak interaction force reflects the short range elastic interactions between the topological loops,
- quantum physics reflects the microscopic perturbations of the lattice expansion field associated to the topological loops, corresponding in fact to perturbations of the scalar gravitational field,
- some microscopic axial properties of a "colored" cubic 3D-lattice could be responsible for the zoology of elementary particles of the standard model,
- the strong interaction force could reflect "colored" stacking faults acting between the topological loops

representing the quarks in the "colored" cubic 3D-lattice,

- the cosmological lattice can expand or contract in the absolute space-time frame, which leads to simple explanations for several still unexplained phenomena of modern cosmology and for the intriguing "dark energy",

- a scalar curvature charge appears in the case of the edge dislocation loops, which does not exist in general relativity, in quantum physics or in the standard model, leading to an explanation for the small asymmetry between matter and anti-matter, for the appearance of an antigravity property of the simplest topological loop (the interstitial edge dislocation loop corresponding to the electron neutrino), which at the same time explains the formation of galaxies and the origin of the "dark matter" .

In fact, the theory of the cosmological lattice with its loop topological singularities presented here is not yet completed, as there remain several questions without answers, as for example the exact nature of the «colored» 3D-lattice and its relation to the Higgs field postulated in the standard model, the detailed topological mechanism and the reason for spin values of 1/2 or 1 of the topological loops, and also other unsolved problems.

It appears however that this theory of a *Newtonian "colored" cubic 3D-lattice containing loop topological singularities* is the first and only *(i)* to combine all known physics in a very simple manner, unifying electromagnetism, relativity, gravitation and quantum physics, *(ii)* to give a simple meaning to the local space-time and the quantum behavior of topological singularities, and *(iii)* to propose simple explanations to well-known problems of modern cosmology and of the standard model of elementary particles.

**Acknowledgements:** I would like to thank Gianfranco D'Anna and Daniele Mari for providing valuable input and comments.